\documentclass[a4paper,11pt]{article}

\makeatletter
\gdef\@fpheader{   }
\gdef\@journal{jhep}

\RequirePackage{amsmath}
\RequirePackage{amssymb}
\RequirePackage{epsfig}
\RequirePackage{CJK}
\RequirePackage{graphicx}
\RequirePackage[numbers,sort&compress]{natbib}
\RequirePackage{color}
\RequirePackage[dvipdfmx
,colorlinks=true
,urlcolor=blue
,anchorcolor=blue
,citecolor=blue
,filecolor=blue
,linkcolor=blue
,menucolor=blue
,pagecolor=blue
,linktocpage=true
,pdfproducer=medialab
,pdfa=true
,CJKbookmarks
]{hyperref}

\newif\ifnotoc\notocfalse
\newif\ifemailadd\emailaddfalse
\newif\iftoccontinuous\toccontinuousfalse

\def\@subheader{\@empty}
\def\@keywords{\@empty}
\def\@abstract{\@empty}
\def\@xtum{\@empty}
\def\@dedicated{\@empty}
\def\@arxivnumber{\@empty}
\def\@collaboration{\@empty}
\def\@collaborationImg{\@empty}
\def\@proceeding{\@empty}
\def\@preprint{\@empty}

\newcommand{\subheader}[1]{\gdef\@subheader{#1}}
\newcommand{\keywords}[1]{\if!\@keywords!\gdef\@keywords{#1}\else%
\PackageWarningNoLine{\jname}{Keywords already defined.\MessageBreak Ignoring last definition.}\fi}
\renewcommand{\abstract}[1]{\gdef\@abstract{#1}}
\newcommand{\dedicated}[1]{\gdef\@dedicated{#1}}
\newcommand{\arxivnumber}[1]{\gdef\@arxivnumber{#1}}
\newcommand{\proceeding}[1]{\gdef\@proceeding{#1}}
\newcommand{\xtumfont}[1]{\textsc{#1}}
\newcommand{\correctionref}[3]{\gdef\@xtum{\xtumfont{#1} \href{#2}{#3}}}
\newcommand\jname{JHEP}
\newcommand\acknowledgments{\section*{Acknowledgments}}

\newcommand\preprint[1]{\gdef\@preprint{\hfill #1}}
\newenvironment{proof}[1][Proof]{\noindent\textbf{#1.} }{\ \rule{0.5em}{0.5em}}

\newcommand\note[2][]{%
\if!#1!%
\stepcounter{footnote}\footnotetext{#2}%
\else%
{\renewcommand\thefootnote{#1}%
\footnotetext{#2}}%
\fi}

\newtoks\auth@toks
\renewcommand{\author}[2][]{%
  \if!#1!%
    \auth@toks=\expandafter{\the\auth@toks#2\ }%
  \else
    \auth@toks=\expandafter{\the\auth@toks#2$^{#1}$\ }%
  \fi
}

\newtoks\affil@toks\newif\ifaffil\affilfalse
\newcommand{\affiliation}[2][]{%
\affiltrue
  \if!#1!%
    \affil@toks=\expandafter{\the\affil@toks{\item[]#2}}%
  \else
    \affil@toks=\expandafter{\the\affil@toks{\item[$^{#1}$]#2}}%
  \fi
}

\newtoks\email@toks\newcounter{email@counter}%
\newcommand{\emailAdd}[1]{%
\emailaddtrue%
\ifnum\theemail@counter>0\email@toks=\expandafter{\the\email@toks, \@email{#1}}%
\else\email@toks=\expandafter{\the\email@toks\@email{#1}}%
\fi\stepcounter{email@counter}}
\newcommand{\@email}[1]{\href{mailto:#1}{\tt #1}}

\newcommand*\collaboration[1]{\gdef\@collaboration{#1}}
\newcommand*\collaborationImg[2][]{\gdef\@collaborationImg{#2}}

\newcommand\afterLogoSpace{\smallskip}
\newcommand\afterSubheaderSpace{\vskip3pt plus 2pt minus 1pt}
\newcommand\afterProceedingsSpace{\vskip21pt plus0.4fil minus15pt}
\newcommand\afterTitleSpace{\vskip23pt plus0.06fil minus13pt}
\newcommand\afterRuleSpace{\vskip23pt plus0.06fil minus13pt}
\newcommand\afterCollaborationSpace{\vskip3pt plus 2pt minus 1pt}
\newcommand\afterCollaborationImgSpace{\vskip3pt plus 2pt minus 1pt}
\newcommand\afterAuthorSpace{\vskip5pt plus4pt minus4pt}
\newcommand\afterAffiliationSpace{\vskip3pt plus3pt}
\newcommand\afterEmailSpace{\vskip16pt plus9pt minus10pt\filbreak}
\newcommand\afterXtumSpace{\par\bigskip}
\newcommand\afterAbstractSpace{\vskip16pt plus9pt minus13pt}
\newcommand\afterKeywordsSpace{\vskip16pt plus9pt minus13pt}
\newcommand\afterArxivSpace{\vskip3pt plus0.01fil minus10pt}
\newcommand\afterDedicatedSpace{\vskip0pt plus0.01fil}
\newcommand\afterTocSpace{\bigskip\medskip}
\newcommand\afterTocRuleSpace{\bigskip\bigskip}
\newlength{\affiliationsSep}
\newcommand\beforetochook{\pagestyle{myplain}\pagenumbering{roman}}

\DeclareFixedFont\trfont{OT1}{phv}{b}{sc}{11}

\renewcommand\maketitle{
\pagestyle{empty}
\thispagestyle{titlepage}
\setcounter{page}{0}
\noindent{\small\scshape\@fpheader}\@preprint\par
\afterLogoSpace
\if!\@subheader!\else\noindent{\trfont{\@subheader}}\fi
\afterSubheaderSpace
\if!\@proceeding!\else\noindent{\sc\@proceeding}\fi
\afterProceedingsSpace
{\LARGE\flushleft\sffamily\bfseries\@title\par}
\afterTitleSpace
\hrule height 1.5\p@%
\afterRuleSpace
\if!\@collaboration!\else
{\Large\bfseries\sffamily\raggedright\@collaboration}\par
\afterCollaborationSpace
\fi
\if!\@collaborationImg!\else
{\normalsize\bfseries\sffamily\raggedright\@collaborationImg}\par
\afterCollaborationImgSpace
\fi
{\bfseries\raggedright\sffamily\the\auth@toks\par}
\afterAuthorSpace
\ifaffil\begin{list}{}{%
\setlength{\leftmargin}{0.28cm}%
\setlength{\labelsep}{0pt}%
\setlength{\itemsep}{\affiliationsSep}%
\setlength{\topsep}{-\parskip}}
\itshape\small%
\the\affil@toks
\end{list}\fi
\afterAffiliationSpace
\ifemailadd 
\noindent\hspace{0.28cm}\begin{minipage}[l]{.9\textwidth}
\begin{flushleft}
\textit{E-mail:} \the\email@toks
\end{flushleft}
\end{minipage}
\else 
\PackageWarningNoLine{\jname}{E-mails are missing.\MessageBreak Plese use \protect\emailAdd\space macro to provide e-mails.}
\fi
\afterEmailSpace
\if!\@xtum!\else\noindent{\@xtum}\afterXtumSpace\fi
\if!\@abstract!\else\noindent{\renewcommand\baselinestretch{.9}\textsc{Abstract:}}\ \@abstract\afterAbstractSpace\fi
\if!\@keywords!\else\noindent{\textsc{Keywords:}} \@keywords\afterKeywordsSpace\fi
\if!\@arxivnumber!\else\noindent{\textsc{ArXiv ePrint:}} \href{http://arxiv.org/abs/\@arxivnumber}{\@arxivnumber}\afterArxivSpace\fi
\if!\@dedicated!\else\vbox{\small\it\raggedleft\@dedicated}\afterDedicatedSpace\fi
\ifnotoc\else
\iftoccontinuous\else\newpage\fi
\beforetochook\hrule
\tableofcontents
\afterTocSpace
\hrule
\afterTocRuleSpace
\fi
\setcounter{footnote}{0}
\pagestyle{myplain}\pagenumbering{arabic}
} 

\renewcommand{\baselinestretch}{1.1}\normalsize
\divide\@tempdima\baselineskip
\@tempcnta=\@tempdima
\skip\@mpfootins = \skip\footins
\renewcommand{\@dotsep}{10000}

\newcommand\ps@myplain{
\pagenumbering{arabic}
\renewcommand\@oddfoot{\hfill-- \thepage\ --\hfill}
\renewcommand\@oddhead{}}
\let\ps@plain=\ps@myplain

\newcommand\ps@titlepage{\renewcommand\@oddfoot{}\renewcommand\@oddhead{}}

\numberwithin{equation}{section}

\renewcommand\section{\@startsection{section}{1}{\z@}%
                                   {-3.5ex \@plus -1.3ex \@minus -.7ex}%
                                   {2.3ex \@plus.4ex \@minus .4ex}%
                                   {\normalfont\large\bfseries}}
\renewcommand\subsection{\@startsection{subsection}{2}{\z@}%
                                   {-2.3ex\@plus -1ex \@minus -.5ex}%
                                   {1.2ex \@plus .3ex \@minus .3ex}%
                                   {\normalfont\normalsize\bfseries}}
\renewcommand\subsubsection{\@startsection{subsubsection}{3}{\z@}%
                                   {-2.3ex\@plus -1ex \@minus -.5ex}%
                                   {1ex \@plus .2ex \@minus .2ex}%
                                   {\normalfont\normalsize\bfseries}}
\renewcommand\paragraph{\@startsection{paragraph}{4}{\z@}%
                                   {1.75ex \@plus1ex \@minus.2ex}%
                                   {-1em}%
                                   {\normalfont\normalsize\bfseries}}
\renewcommand\subparagraph{\@startsection{subparagraph}{5}{\parindent}%
                                   {1.75ex \@plus1ex \@minus .2ex}%
                                   {-1em}%
                                   {\normalfont\normalsize\bfseries}}

\def\fnum@figure{\textbf{\figurename\nobreakspace\thefigure}}
\def\fnum@table{\textbf{\tablename\nobreakspace\thetable}}

\long\def\@makecaption#1#2{%
  \vskip\abovecaptionskip
  \sbox\@tempboxa{\small #1. #2}%
  \ifdim \wd\@tempboxa >\hsize
    \small #1. #2\par
  \else
    \global \@minipagefalse
    \hb@xt@\hsize{\hfil\box\@tempboxa\hfil}%
  \fi
  \vskip\belowcaptionskip}

\renewenvironment{thebibliography}[1]{%
\begin{oldthebibliography}{#1}%
\small%
\raggedright%
\setlength{\itemsep}{5pt plus 0.2ex minus 0.05ex}%
}%
{%
\end{oldthebibliography}%
}

\makeatother

\usepackage[T1]{fontenc} 
\usepackage{romannum} 
\usepackage{CJK}

\title{{\boldmath Acoustic scattering theory without large-distance asymptotics}}

\author[a]{Chi-Chun Zhou,}
\author[b,a]{Wen-Du Li,}
\author[a*]{and Wu-Sheng Dai}\note{daiwusheng@tju.edu.cn.}

\affiliation[a]{Department of Physics, Tianjin University, Tianjin 300350, P.R. China}
\affiliation[b]{Theoretical Physics Division, Chern Institute of Mathematics, Nankai University, Tianjin, 300071, P. R. China}

\abstract{In conventional acoustic scattering theory, a large-distance asymptotic
approximation is employed. In this approximation, a far-field pattern, an
asymptotic approximation of the exact result, is used to describe a scattering
process. The information of the distance between the target and the observer,
however, is lost in the large-distance asymptotic approximation. In this
paper, we provide a rigorous theory of acoustic scattering without the
large-distance asymptotic approximation. The acoustic scattering
treatment\ developed in this paper provides an improved description for the
acoustic wave outside the target. Moreover, as examples, we consider acoustic
scattering on a rigid sphere and on a nonrigid sphere. We also illustrate the
influence of the near target effect on the angular distribution of outgoing
waves. It is shown that for long wavelength acoustic scattering, the near
target effect must be reckoned in.
}

\keywords{Acoustic scattering; Large-distance asymptotics; Near target effect.}

\begin{document} 
\begin{CJK*}{GBK}{song}
\maketitle 

\flushbottom

\section{Introduction}

Solving an acoustic scattering problem boils down to solving a Helmholtz
equation
\begin{equation}
\left(  \mathbf{\nabla}^{2}+k^{2}\right)  u\left(  \mathbf{x}\right)  =0
\label{2222}%
\end{equation}
with a given boundary condition, where $k$ is the wave number and $u\left(
\mathbf{x}\right)  $ describes the complex amplitude of acoustic pressures
\cite{morse1968theoretical,pike2001scattering,kress2013inverse,schroeder2007handbook}%
. If the incident wave is a plane wave, the solution of Eq. (\ref{2222}) can
be written as%
\begin{equation}
u\left(  \mathbf{x}\right)  =e^{i\mathbf{k\cdot x}}+u^{s}\left(
\mathbf{x}\right)  , \label{oko}%
\end{equation}
where $\mathbf{k}$ is the wave vector of the incident wave and$\ u^{s}\left(
\mathbf{x}\right)  $ is the outgoing wave
\cite{kress2013inverse,pike2001scattering,schroeder2007handbook}. Then the
task of solving $u\left(  \mathbf{x}\right)  $ is converted into solving the
outgoing wave $u^{s}\left(  \mathbf{x}\right)  $.

In conventional acoustic scattering theory, the outgoing wave $u^{s}\left(
\mathbf{x}\right)  $ is approximately solved under a large-distance
asymptotics \cite{kress2013inverse,pike2001scattering,schroeder2007handbook}:
\begin{equation}
u^{s}\left(  \mathbf{x}\right)  =\frac{e^{ik\left\vert \mathbf{x}\right\vert
}}{\left\vert \mathbf{x}\right\vert }u_{\infty}\left(  \frac{\mathbf{x}%
}{\left\vert \mathbf{x}\right\vert }\right)  +O\left(  \frac{1}{\left\vert
\mathbf{x}\right\vert ^{2}}\right)  , \label{9988}%
\end{equation}
where $u_{\infty}\left(  \frac{\mathbf{x}}{\left\vert \mathbf{x}\right\vert
}\right)  $ is the far-field pattern of the outgoing wave. Then, instead of
the exact outgoing wave $u^{s}\left(  \mathbf{x}\right)  $, one turns to
calculate the far-field pattern $u_{\infty}\left(  \frac{\mathbf{x}%
}{\left\vert \mathbf{x}\right\vert }\right)  $.

The large-distance asymptotic approximation supposes that the distance between
the target and the observer is infinite. Obviously, such an approximation
loses the information of the distance. As a result, the observable quantity,
such as the scattering intensity $I\left(  \mathbf{x}\right)  $, depends only
on the angle of emergence.

In realistic acoustic scattering, however, the wavelength of acoustic waves is
often comparable with the distance between the target and the observer. In
long wavelength acoustic-wave scattering, the large-distance asymptotic
approximation is not accurate enough and a scattering theory without
large-distance asymptotics is needed.

Recently, a rigorous scattering theory without large-distance asymptotics in
quantum mechanics is developed \cite{liu2014scattering,li2016scattering}. In
this scattering theory, the information of the distance between the target and
the observer is taken into account, since there is no large-distance
asymptotic approximation. Taking advantage of this rigorous scattering theory,
in this\textbf{ }paper, we develop a rigorous acoustic scattering
theory\textbf{ }without large-distance asymptotic approximation. The acoustic
scattering theory developed in the present paper is valid outside the target
since this treatment depends on the assumption that the influence of the
target must vanish at large distance.

Acoustic scattering deals with the propagation of sounds. There are many
researches on acoustic scattering, for example, acoustic waves scattering on
different targets such as fluid spheroids \cite{Gonz2016A} and finite rigid
plates \cite{Ayton2016Acoustic}. Underwater acoustic scattering is also an
important issue, such as underwater acoustic scattering signal separation
\cite{Jia2017Rigid}, acoustic scattering from underwater elastic objects
\cite{Chai2017Application}, acoustic scattering by suspended flocculating
sediments \cite{Thorne2014Modelling} and acoustic scattering in gassy soft
marine sediments \cite{Mantouka2016Modelling}. Other problems such as acoustic
scattering-resonance \cite{Steinbach2017Combined}, acoustic scattering
reductions \cite{Dutrion2016Acoustic}, acoustic scattering in nonuniform
potential flows \cite{Karimi2017Acoustic}, designing the bilaminate acoustic
cloak \cite{Guild2014Cloaking}, control of acoustic absorptions
\cite{Merkel2015Control}, acoustic interaction forces \cite{silva2014acoustic}%
, acoustic radiation forces \cite{Mitri2015Acoustic}, and scattering effects
on an acoustic black hole \cite{Denis2016Scattering} are also considered.

In section \ref{Helmholtz}, we construct an rigorous acoustic scattering
theory\textbf{ }without large-distance asymptotics. In sections
\ref{rigid sphere} and\ \ref{homogeneous sphere}, we consider two acoustic
scattering processes, plane waves scattering on a rigid sphere and on a
nonrigid sphere. The conclusion is summarized in section \ref{Conclusion}.

\section{Acoustic scattering theory without large-distance asymptotics
\label{Helmholtz}}

As mentioned above, in conventional acoustic scattering theory, instead of the
outgoing wave $u^{s}\left(  \mathbf{x}\right)  $, one turns to find the
far-field pattern $u_{\infty}\left(  \frac{\mathbf{x}}{\left\vert
\mathbf{x}\right\vert }\right)  $, the large-distance asymptotic approximation
of $u^{s}\left(  \mathbf{x}\right)  $. In this section, instead of the
far-field pattern $u_{\infty}\left(  \frac{\mathbf{x}}{\left\vert
\mathbf{x}\right\vert }\right)  $, we calculate the exact result of the
outgoing wave $u^{s}\left(  \mathbf{x}\right)  $ by introducing an exact
pattern $u_{exact}\left(  \mathbf{x}\right)  $ which is an exact version of
the far-field pattern $u_{\infty}\left(  \frac{\mathbf{x}}{\left\vert
\mathbf{x}\right\vert }\right)  $.

\subsection{Exact pattern}

Introduce the exact pattern $u_{exact}\left(  \mathbf{x}\right)  $ by
rewriting the outgoing wave $u^{s}\left(  \mathbf{x}\right)  $ in Eq.
(\ref{oko}) as%
\begin{equation}
u^{s}\left(  \mathbf{x}\right)  =\frac{e^{ik\left\vert \mathbf{x}\right\vert
}}{\left\vert \mathbf{x}\right\vert }u_{exact}\left(  \mathbf{x}\right)  .
\label{1113}%
\end{equation}

As will be shown later, the exact pattern $u_{exact}\left(  \mathbf{x}\right)
$ recovers the far-field pattern as $\left\vert \mathbf{x}\right\vert $ goes
to infinity:
\begin{equation}
u_{exact}\left(  \mathbf{x}\right)  \overset{\left\vert \mathbf{x}\right\vert
\rightarrow\infty}{\sim}u_{\infty}\left(  \frac{\mathbf{x}}{\left\vert
\mathbf{x}\right\vert }\right)  .
\end{equation}

Now we represent the scattering intensity by the exact pattern $u_{exact}%
\left(  \mathbf{x}\right)  $.

\textit{The scattering intensity }$I\left(  \mathbf{x}\right)  $\textit{ can
be represented in terms of the exact pattern }$u_{exact}\left(  \mathbf{x}%
\right)  $\textit{ as }%
\begin{equation}
I\left(  \mathbf{x}\right)  =\frac{\rho_{0}\omega k}{\left\vert \mathbf{x}%
\right\vert ^{2}}\left\vert u_{exact}\left(  \mathbf{x}\right)  \right\vert
^{2}, \label{55}%
\end{equation}
\textit{where }$\rho_{0}$\textit{ is the density of the medium, }$\omega
$\textit{ is the frequency of the acoustic wave, and }$k=2\pi/\lambda$\textit{
with }$\lambda$\textit{ the wavelength.}

\begin{proof}
The scattering intensity $I\left(  \mathbf{x}\right)  $ is defined as the
component of the energy flow in the radial direction,
\begin{equation}
I\left(  \mathbf{x}\right)  =\mathbf{J}_{s}\left(  \mathbf{x}\right)
\cdot\mathbf{e}_{r}, \label{56}%
\end{equation}
where the energy flow $\mathbf{J}\left(  \mathbf{x}\right)  $ is defined as
\cite{pike2001scattering,morse1968theoretical}
\begin{equation}
\mathbf{J}\left(  \mathbf{x}\right)  =\frac{1}{2}\frac{i}{\omega\rho_{0}%
}\left[  u\left(  \mathbf{x}\right)  \mathbf{\nabla}u^{\ast}\left(
\mathbf{x}\right)  -u^{\ast}\left(  \mathbf{x}\right)  \mathbf{\nabla}u\left(
\mathbf{x}\right)  \right]  . \label{11}%
\end{equation}
Substituting Eq. (\ref{1113}) into Eqs. (\ref{11}) and (\ref{56}) proves Eq.
(\ref{55}).
\end{proof}

The result obtained in the present paper is based the short-range interaction
scattering theory. The short-range scattering theory is based on the
assumption that the interaction must decrease fast enough. The acoustic
scattering in which the influence of the target is limited to a given spatial
scale, i.e., the influence of the target decreases to zero outside the target,
is a short-range interaction. For such a kind of scattering, the result given
above is valid outside the target.

\subsection{Scattering phase shift}

In acoustic scattering, the target plays the role of the potential in
quantum-mechanical scattering. In this section, we introduce the scattering
phase shift $\delta_{l}$ and express the exact pattern $u_{exact}\left(
\mathbf{x}\right)  $ in terms of the scattering phase shift.

\textit{The phase shift }$\delta_{l}$\textit{. }For scattering of a plane wave
by a spherically symmetric target, the solution of Eq. (\ref{2222}) can be
written as a linear combination of the spherical Hankel function
\cite{kress2013inverse,li2016scattering}%
\begin{equation}
u\left(  \mathbf{x}\right)  =\sum_{l=0}^{\infty}C_{l}h_{l}^{\left(  2\right)
}\left(  k\left\vert \mathbf{x}\right\vert \right)  P_{l}\left(  \cos
\theta\right)  +\sum_{l=0}^{\infty}D_{l}h_{l}^{\left(  1\right)  }\left(
k\left\vert \mathbf{x}\right\vert \right)  P_{l}\left(  \cos\theta\right)
,\label{7878}%
\end{equation}
where $h_{l}^{\left(  1\right)  }\left(  x\right)  $ and $h_{l}^{\left(
2\right)  }\left(  x\right)  $ are the spherical\textbf{ }Hankel functions of
first and second kinds and $P_{l}\left(  x\right)  $ is the Legendre function
\cite{olver2010nist}, $C_{l}$ and $D_{l}$ are coefficients, and $\theta$ is
the intersection angle between the direction of incident plane wave and the radial.

The scattering phase shift $\delta_{l}$ is introduced as
\cite{li2016scattering}
\begin{equation}
e^{2i\delta_{l}}=\frac{D_{l}}{C_{l}}. \label{88888}%
\end{equation}

\textit{The exact pattern }$u_{exact}\left(  \mathbf{x}\right)  $\textit{.
}Now we express the exact pattern $u_{exact}\left(  \mathbf{x}\right)  $
explicitly in terms of the scattering phase shift $\delta_{l}$.

\textit{The exact pattern }$u_{exact}\left(  \mathbf{x}\right)  $\textit{ can
be represented in terms of the phase shift }$\delta_{l}$\textit{ as}%
\begin{equation}
u_{exact}\left(  \mathbf{x}\right)  =\frac{1}{2ik}\sum_{l=0}^{\infty}\left(
2l+1\right)  \left(  e^{2i\delta_{l}}-1\right)  P_{l}\left(  \cos
\theta\right)  y_{l}\left(  -\frac{1}{ik\left\vert \mathbf{x}\right\vert
}\right)  , \label{66}%
\end{equation}
\textit{where }$y_{l}\left(  x\right)  =\sum_{k=0}^{l}\frac{\left(
l+k\right)  !}{k!\left(  l-k\right)  !}\left(  \frac{x}{2}\right)  ^{k}%
$\textit{ is the Bessel polynomial.}

\begin{proof}
By the expression of the outgoing wave Eq. (\ref{1113}), the solution of the
Helmholtz equation Eq. (\ref{oko}) can be written as%
\begin{equation}
u\left(  \mathbf{x}\right)  =e^{ik\left\vert \mathbf{x}\right\vert \cos\theta
}+\frac{e^{ik\left\vert \mathbf{x}\right\vert }}{\left\vert \mathbf{x}%
\right\vert }u_{exact}\left(  \mathbf{x}\right)  . \label{9879}%
\end{equation}
The plane wave $e^{ik\left\vert \mathbf{x}\right\vert \cos\theta}$ in Eq.
(\ref{9879}) can be exactly expanded as \cite{li2016scattering}
\begin{equation}
e^{ik\left\vert \mathbf{x}\right\vert \cos\theta}=\sum_{l=0}^{\infty}\left(
2l+1\right)  i^{l}M_{l}\left(  -\frac{1}{ik\left\vert \mathbf{x}\right\vert
}\right)  \frac{1}{k\left\vert \mathbf{x}\right\vert }\sin\left[  k\left\vert
\mathbf{x}\right\vert -\frac{l\pi}{2}+\Delta_{l}\left(  -\frac{1}{ik\left\vert
\mathbf{x}\right\vert }\right)  \right]  P_{l}\left(  \cos\theta\right)  ,
\label{expan}%
\end{equation}
where $M_{l}\left(  x\right)  =\left\vert y_{l}\left(  x\right)  \right\vert $
and $\Delta_{l}\left(  x\right)  =\arg y_{l}\left(  x\right)  $ are the
modulus and argument of the Bessel polynomial $y_{l}\left(  x\right)  $.
Substituting Eq. (\ref{expan}) into Eq. (\ref{9879}) gives
\begin{equation}
u\left(  \mathbf{x}\right)  =\sum_{l=0}^{\infty}\left(  2l+1\right)
i^{l}M_{l}\left(  -\frac{1}{ik\left\vert \mathbf{x}\right\vert }\right)
\frac{1}{k\left\vert \mathbf{x}\right\vert }\sin\left[  k\left\vert
\mathbf{x}\right\vert -\frac{l\pi}{2}+\Delta_{l}\left(  -\frac{1}{ik\left\vert
\mathbf{x}\right\vert }\right)  \right]  P_{l}\left(  \cos\theta\right)
+\frac{e^{ik\left\vert \mathbf{x}\right\vert }}{\left\vert \mathbf{x}%
\right\vert }\sum_{l=0}^{\infty}u_{exact}^{\left(  l\right)  }\left(
\mathbf{x}\right)  , \label{333}%
\end{equation}
where $u_{exact}\left(  \mathbf{x}\right)  $ is expressed as%
\begin{equation}
u_{exact}\left(  \mathbf{x}\right)  =\sum_{l=0}^{\infty}u_{exact}^{\left(
l\right)  }\left(  \mathbf{x}\right)  . \label{5556}%
\end{equation}

Eq. (\ref{7878}) can be expressed as \cite{li2016scattering}
\begin{equation}
u\left(  \mathbf{x}\right)  =\sum_{l=0}^{\infty}A_{l}M_{l}\left(  -\frac
{1}{ik\left\vert \mathbf{x}\right\vert }\right)  \frac{1}{k\left\vert
\mathbf{x}\right\vert }\sin\left[  k\left\vert \mathbf{x}\right\vert
-\frac{l\pi}{2}+\delta_{l}+\Delta_{l}\left(  -\frac{1}{ik\left\vert
\mathbf{x}\right\vert }\right)  \right]  P_{l}\left(  \cos\theta\right)  ,
\label{4545}%
\end{equation}
where $A_{l}=2\sqrt{C_{l}D_{l}}.$

Rewriting the trigonometric function in Eqs. (\ref{333}) and (\ref{4545}) in
terms of the exponential function and equalling the coefficients of $e^{\pm
ik\left\vert \mathbf{x}\right\vert }$ give
\begin{equation}
A_{l}=\frac{1}{2}i^{l}\left(  2l+1\right)  e^{i\delta_{l}} \label{66666}%
\end{equation}
and
\begin{equation}
u_{exact}^{\left(  l\right)  }\left(  \mathbf{x}\right)  =\frac{1}{2ik}\left(
2l+1\right)  \left(  e^{2i\delta_{l}}-1\right)  P_{l}\left(  \cos
\theta\right)  y_{l}\left(  -\frac{1}{ik\left\vert \mathbf{x}\right\vert
}\right)  . \label{888888}%
\end{equation}
Substituting Eq. (\ref{888888}) into Eq. (\ref{5556}) proves Eq. (\ref{66}).
\end{proof}

\subsection{Scattering intensity and sound pressure}

The scattering intensity $I\left(  \mathbf{x}\right)  $ in Eq. (\ref{55}) is
represented by the exact pattern $u_{exact}\left(  \mathbf{x}\right)  $ and
the exact pattern $u_{exact}\left(  \mathbf{x}\right)  $ in Eq. (\ref{66}) is
represented by the phase shift $\delta_{l}$. In this section, we represent the
scattering intensity $I\left(  \mathbf{x}\right)  $ and the pressure $p\left(
\mathbf{x},t\right)  $ in terms of the phase shift $\delta_{l}$.

The scattering intensity can be obtained by substituting Eq. (\ref{66}) into
Eq. (\ref{55}):%
\begin{equation}
I\left(  \mathbf{x}\right)  =\frac{\rho_{0}\omega k}{\left\vert \mathbf{x}%
\right\vert ^{2}}\left\vert \sum_{l=0}^{\infty}\left(  2l+1\right)  \frac
{1}{2k}\left(  e^{2i\delta_{l}}-1\right)  y_{l}\left(  -\frac{1}{ik\left\vert
\mathbf{x}\right\vert }\right)  P_{l}\left(  \cos\theta\right)  \right\vert
^{2}. \label{77}%
\end{equation}

The sound pressure $p\left(  \mathbf{x},t\right)  $ is%
\begin{equation}
p\left(  \mathbf{x},t\right)  =\operatorname{Re}\left[  u\left(
\mathbf{x}\right)  e^{-i\omega t}\right]  , \label{pp}%
\end{equation}
where $u\left(  \mathbf{x}\right)  $ is given by Eqs. (\ref{4545}) and
(\ref{66666}). Then we have%
\begin{equation}
p\left(  \mathbf{x},t\right)  =\frac{\cos\omega t}{2k\left\vert \mathbf{x}%
\right\vert }\sum_{l=0}^{\infty}M_{l}\left(  -\frac{1}{ik\left\vert
\mathbf{x}\right\vert }\right)  i^{l}\left(  2l+1\right)  e^{i\delta_{l}}%
\sin\left[  k\left\vert \mathbf{x}\right\vert -\frac{l\pi}{2}+\delta
_{l}+\Delta_{l}\left(  -\frac{1}{ik\left\vert \mathbf{x}\right\vert }\right)
\right]  P_{l}\left(  \cos\theta\right)  .
\end{equation}

\section{Acoustic scattering on rigid sphere \label{rigid sphere}}

In this section, as an example, we consider the problem of a plane wave
scattering on a rigid sphere.

\subsection{Scattering intensity}

The phase shift $\delta_{l}$ of a plane wave scattered by a rigid sphere can
be obtained from the boundary condition. The boundary condition of a rigid
sphere is \cite{schroeder2007handbook,morse1968theoretical}
\begin{equation}
\left.  \frac{1}{i\omega\rho_{0}}\mathbf{\nabla}u\left(  \mathbf{x}\right)
\right\vert _{\left\vert \mathbf{x}\right\vert =a}=0, \label{99}%
\end{equation}
where $a$ is the radius of the sphere. Substituting Eq. (\ref{7878}),\textbf{
}$D_{l}=\frac{1}{4}i^{l}\left(  2l+1\right)  e^{2i\delta_{l}}$,\ and
$C_{l}=\frac{1}{4}i^{l}\left(  2l+1\right)  $ \cite{li2016scattering} into the
boundary condition (\ref{99}) gives%
\begin{equation}
e^{2i\delta_{l}}=-\frac{\left.  dh_{l}^{(2)}\left(  kr\right)  /dr\right\vert
_{r=a}}{\left.  dh_{l}^{(1)}\left(  kr\right)  /dr\right\vert _{r=a}}%
=\frac{\frac{l}{ka}h_{l}^{\left(  2\right)  }\left(  ka\right)  -h_{l+1}%
^{\left(  2\right)  }\left(  ka\right)  }{\frac{l}{ka}h_{l}^{\left(  1\right)
}\left(  ka\right)  -h_{l+1}^{\left(  1\right)  }\left(  ka\right)  },
\label{00001}%
\end{equation}
where $\frac{d}{dz}h_{\nu}^{\left(  1,2\right)  }\left(  z\right)  =\frac{\nu
}{z}h_{\nu}^{\left(  1,2\right)  }\left(  z\right)  -h_{\nu+1}^{\left(
1,2\right)  }\left(  z\right)  $ is used \cite{olver2010nist}. The phase
shift, by noting that $h_{l}^{\left(  2\right)  \ast}\left(  z\right)
=h_{l}^{\left(  1\right)  }\left(  z\right)  $, then reads
\begin{equation}
\delta_{l}=\arg\left(  \frac{l}{ka}h_{l}^{\left(  2\right)  }\left(
ka\right)  -h_{l+1}^{\left(  2\right)  }\left(  ka\right)  \right)  .
\end{equation}

The scattering intensity then can be achieved by substituting Eq.
(\ref{00001}) into Eq. (\ref{77}):%
\begin{equation}
I\left(  \mathbf{x}\right)  =\frac{\rho_{0}\omega k}{\left\vert \mathbf{x}%
\right\vert ^{2}}\left\vert \sum_{l=0}^{\infty}\left(  2l+1\right)  \frac
{1}{2k}\left[  \exp\left(  2i\arg\left(  \frac{l}{ka}h_{l}^{\left(  2\right)
}\left(  ka\right)  -h_{l+1}^{\left(  2\right)  }\left(  ka\right)  \right)
\right)  +1\right]  y_{l}\left(  -\frac{1}{ik\left\vert \mathbf{x}\right\vert
}\right)  P_{l}(\cos\theta)\right\vert ^{2}. \label{100}%
\end{equation}

\textit{Rayleigh scattering.} Taking only the $s$-wave and $p$-wave
contributions into account and assuming that $ka\ll1$, we arrive at
\begin{equation}
I\left(  \mathbf{x}\right)  \sim\frac{\rho_{0}\omega k^{5}a^{6}}{9\left\vert
\mathbf{x}\right\vert ^{2}}\left(  1-\frac{3}{2}\cos\theta\right)  ^{2}\left[
1+\frac{1}{k^{2}\left\vert \mathbf{x}\right\vert ^{2}}\left(  \frac
{3\cos\theta}{2-3\cos\theta}\right)  ^{2}\right]  , \label{101}%
\end{equation}
where $y_{0}\left(  -\frac{1}{ik\left\vert \mathbf{x}\right\vert }\right)  =1$
and $y_{1}\left(  -\frac{1}{ik\left\vert \mathbf{x}\right\vert }\right)
=1-\frac{1}{ik\left\vert \mathbf{x}\right\vert }$ are used. One can see that
the scattering intensity $I\left(  \mathbf{x}\right)  $ is a function of both
the angle $\theta$ and the distance $\left\vert \mathbf{x}\right\vert $. When
$\left\vert \mathbf{x}\right\vert \rightarrow\infty$, $I\left(  \mathbf{x}%
\right)  $ reduces to the result of the large-distance asymptotic
approximation in conventional acoustic scattering theory
\cite{schroeder2007handbook,morse1968theoretical}:%
\begin{equation}
I_{\infty}\left(  \mathbf{x}\right)  \sim\frac{\rho_{0}\omega k^{5}a^{6}%
}{9\left\vert \mathbf{x}\right\vert ^{2}}\left(  1-\frac{3}{2}\cos
\theta\right)  ^{2}. \label{102}%
\end{equation}

\subsection{Near target effect}

The large-distance asymptotic approximation in conventional scattering theory
losses the information of the distance between the target and the observer. In
the following, we show the influence of the near target effect in scattering
cross sections.

\begin{figure}[ptb]
\centering
\includegraphics[width=0.8\textwidth]{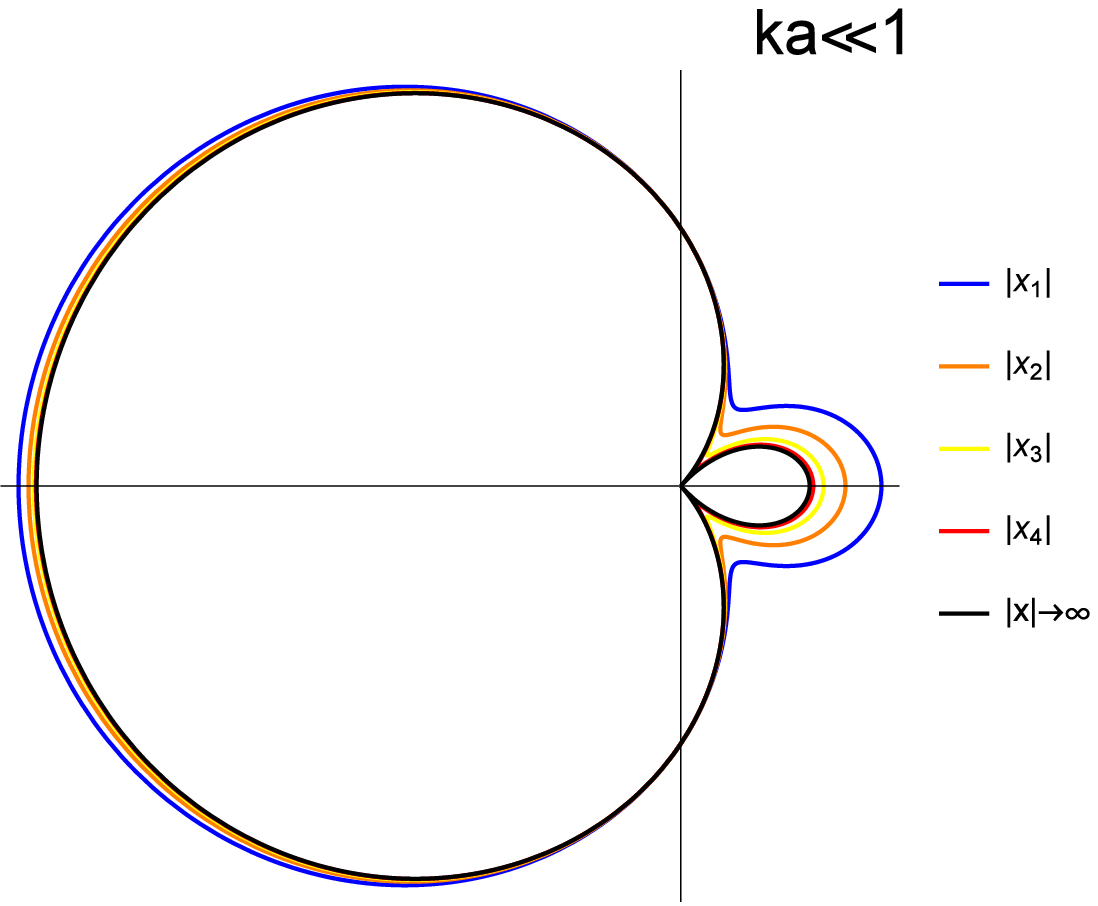}\newline\caption{A sketch
of the angular distribution of a plane wave scattered by a rigid sphere of
radius $a$: $\left(  \frac{d\sigma}{d\Omega}\right)  ^{1/2}/a$ versus the
polar angle $\theta$ for $ka\ll1$ up to the $f$-wave contribution at different
distances $\left\vert x_{1}\right\vert <\left\vert x_{2}\right\vert
<\left\vert x_{3}\right\vert <\left\vert x_{4}\right\vert <\left\vert
x\right\vert $.}%
\label{ka_0}%
\end{figure}

\begin{figure}[ptb]
\centering
\includegraphics[width=0.8\textwidth]{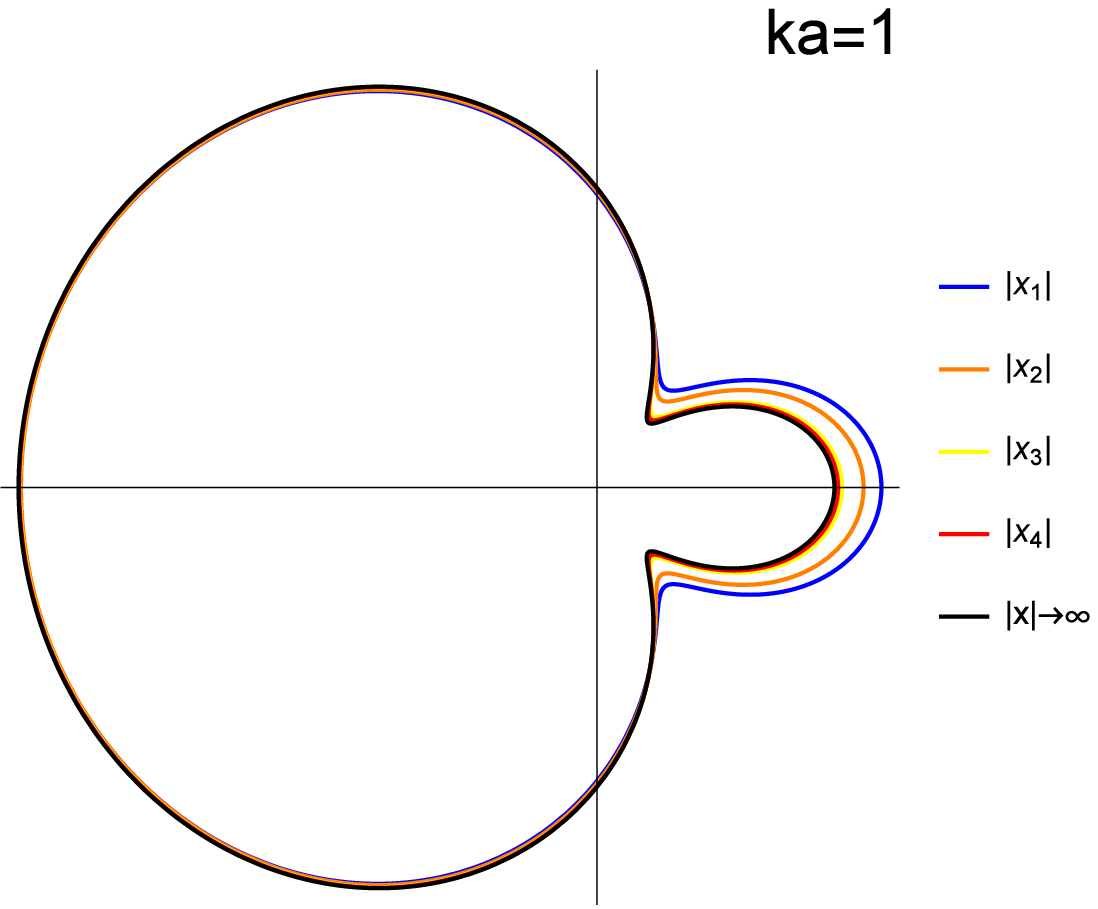}\newline\caption{A sketch
of the angular distribution of a plane wave scattered by a rigid sphere of
radius $a$: $\left(  \frac{d\sigma}{d\Omega}\right)  ^{1/2}/a$ versus the
polar angle $\theta$ for $ka=1$ up to the $f$-wave contribution at different
distances $\left\vert x_{1}\right\vert <\left\vert x_{2}\right\vert
<\left\vert x_{3}\right\vert <\left\vert x_{4}\right\vert <\left\vert
x\right\vert $.}%
\label{ka_1}%
\end{figure}

\begin{figure}[ptb]
\centering
\includegraphics[width=0.8\textwidth]{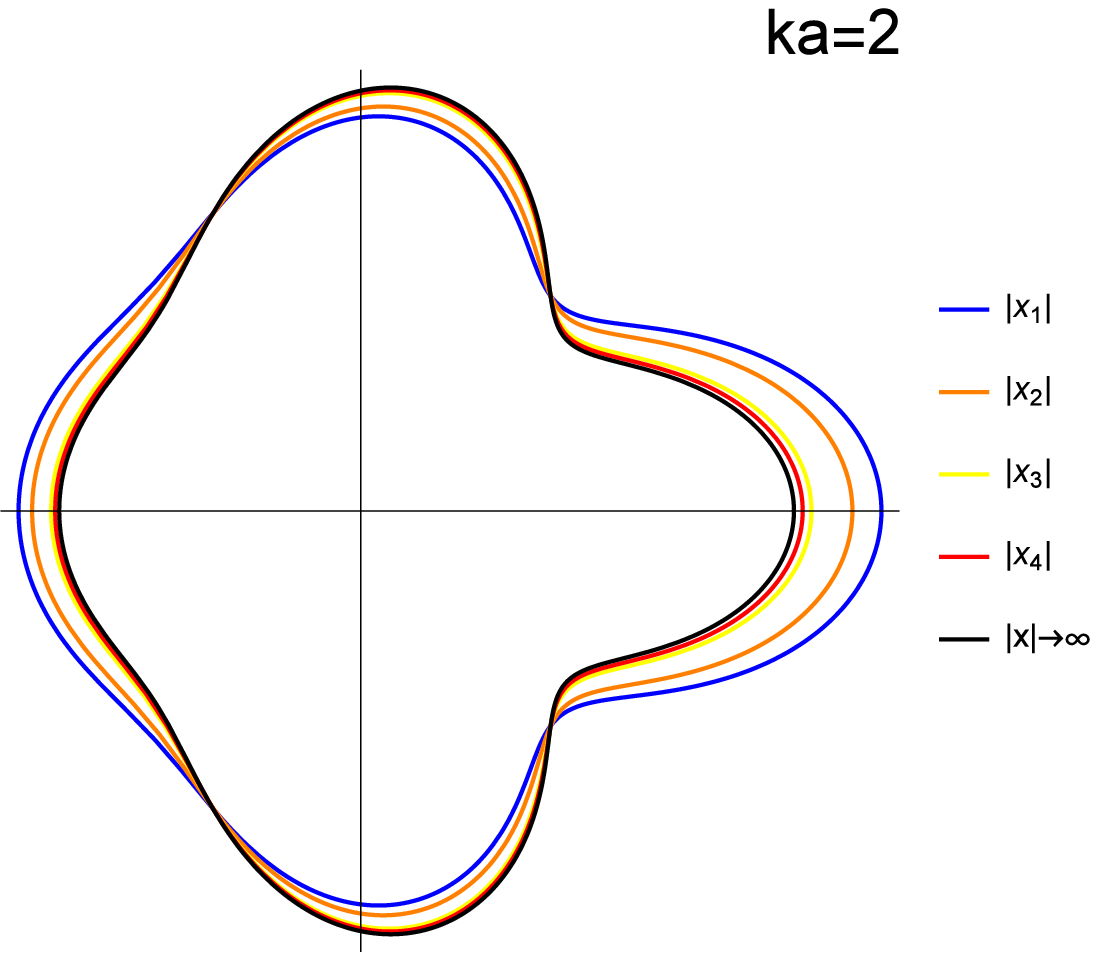}\newline\caption{A sketch
of the angular distribution of a plane wave scattered by a rigid sphere of
radius $a$: $\left(  \frac{d\sigma}{d\Omega}\right)  ^{1/2}/a$ versus the
polar angle $\theta$ for $ka=2$ up to the $f$-wave contribution at different
distances $\left\vert x_{1}\right\vert <\left\vert x_{2}\right\vert
<\left\vert x_{3}\right\vert <\left\vert x_{4}\right\vert <\left\vert
x\right\vert $.}%
\label{ka_2}%
\end{figure}

\begin{figure}[ptb]
\centering
\includegraphics[width=0.8\textwidth]{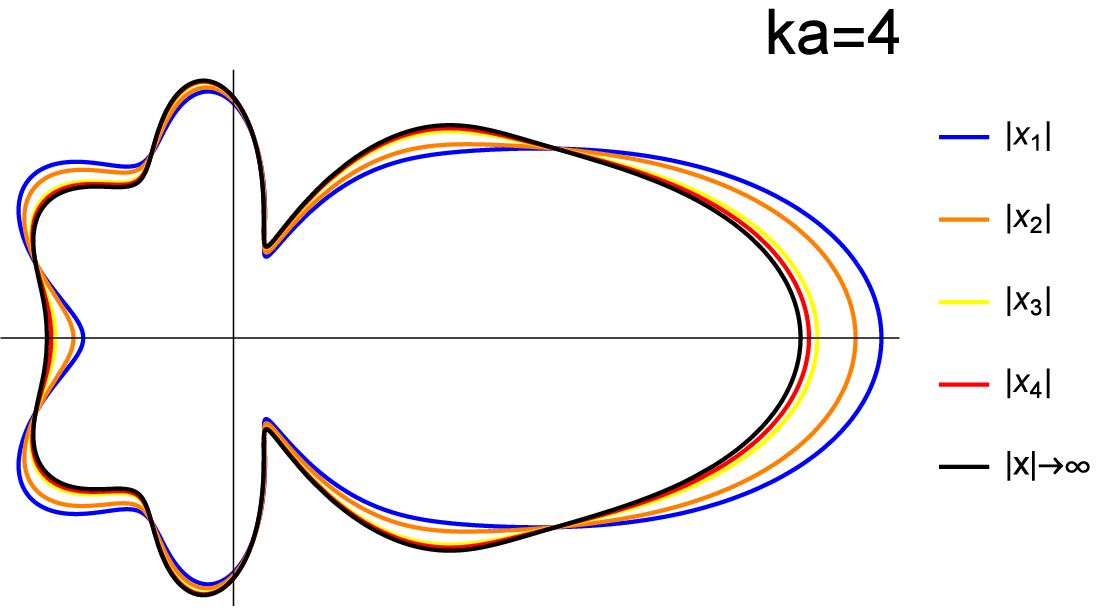}\newline\caption{A sketch
of the angular distribution of a plane wave scattered by a rigid sphere of
radius $a$: $\left(  \frac{d\sigma}{d\Omega}\right)  ^{1/2}/a$ versus the
polar angle $\theta$ for $ka=4$ up to the $f$-wave contribution at different
distances $\left\vert x_{1}\right\vert <\left\vert x_{2}\right\vert
<\left\vert x_{3}\right\vert <\left\vert x_{4}\right\vert <\left\vert
x\right\vert $.}%
\label{ka_4}%
\end{figure}

The differential scattering cross section is given by
\cite{schroeder2007handbook}
\begin{equation}
\frac{d\sigma}{d\Omega}=I\left(  \mathbf{x}\right)  \frac{\left\vert
\mathbf{x}\right\vert ^{2}}{\rho_{0}\omega k}, \label{jiemian}%
\end{equation}
where the scattering intensity $I\left(  \mathbf{x}\right)  $ is given by Eq.
(\ref{100}). In figures (\ref{ka_0}) to (\ref{ka_4}), we illustrate $\left(
\frac{d\sigma}{d\Omega}\right)  ^{1/2}/a$ versus the polar angel $\theta$ at
different distance $\left\vert x\right\vert $ for different wave lengths up to
the $f$-wave contribution.

From the figures we can see that the near target effect can not be ignored
when the distance is comparable with the wave length. The exact result
recovers the large-distance asymptotics in conventional scattering theory as
the distance goes to infinity.

In order to show the near target effect, we consider the deviation
$\Delta\sigma=\sigma-\sigma_{\infty}$. By Eqs. (\ref{jiemian}), (\ref{101}),
and (\ref{102}), we have
\begin{equation}
\frac{d\Delta\sigma}{d\Omega}=\frac{k^{2}a^{6}}{4\left\vert \mathbf{x}%
\right\vert ^{2}}\cos^{2}\theta.
\end{equation}
where we take only the $s$-wave and $p$-wave contributions into account and
assume that $ka\ll1$.

\subsection{Power}

The power of an acoustic scattering on a sphere can be calculated by
\cite{schroeder2007handbook,morse1968theoretical}%
\begin{equation}
P_{sc}=\int_{0}^{2\pi}\int_{0}^{\pi}I\left(  \mathbf{x}\right)  \left\vert
\mathbf{x}\right\vert ^{2}\sin\theta d\theta d\phi. \label{10222}%
\end{equation}
Substituting Eq. (\ref{101}) into Eq. (\ref{10222}) gives
\begin{equation}
P_{sc}\left(  \omega,\mathbf{x}\right)  \sim\frac{7\pi\rho_{0}\omega
a^{6}k^{5}}{9}\left(  1+\frac{3}{7k^{2}\left\vert \mathbf{x}\right\vert ^{2}%
}\right)  . \label{1010}%
\end{equation}

The large-distance asymptotics then reads
\begin{equation}
P_{sc\infty}\left(  \omega\right)  \sim\frac{7\pi\rho_{0}\omega a^{6}k^{5}}%
{9}. \label{1020}%
\end{equation}
This is just the result given by the conventional scattering theory
\cite{schroeder2007handbook}.

\section{Acoustic scattering on nonrigid sphere \label{homogeneous sphere}}

In this section, we consider the problem of a plane wave scattering on a
nonrigid sphere.

\subsection{Scattering intensity}

For a plane wave scattering on a nonrigid sphere, the phase shift $\delta_{l}$
can be obtained from the boundary condition of a nonrigid sphere
\cite{schroeder2007handbook,morse1968theoretical}.

Inside a nonrigid sphere, the solution of the Helmholtz equation can be
written as \cite{schroeder2007handbook,morse1968theoretical}
\begin{equation}
u_{e}\left(  \mathbf{x}\right)  =i\omega\rho_{e}\sum_{l=0}^{\infty}A_{l}%
j_{l}\left(  k_{e}\left\vert \mathbf{x}\right\vert \right)  P_{l}\left(
\cos\theta\right)  ,\label{444}%
\end{equation}
where $k_{e}=\omega/c_{e}$ with $c_{e}$ the velocity of the acoustic wave
inside the nonrigid sphere, $\omega$ the frequency of the acoustic wave,
$A_{l}$ is the coefficient, and $\rho_{e}$ the density of the nonrigid sphere.
By Eq. (\ref{7878}) with\textbf{ }$D_{l}=\frac{1}{4}i^{l}\left(  2l+1\right)
e^{2i\delta_{l}}\ $and $C_{l}=\frac{1}{4}i^{l}\left(  2l+1\right)  $
\cite{li2016scattering}, the solution $u_{o}\left(  \mathbf{x}\right)  $
outside the sphere reads%
\begin{equation}
u_{o}\left(  \mathbf{x}\right)  =\frac{i\omega\rho_{0}}{4}\sum_{l=0}^{\infty
}i^{l}\left(  2l+1\right)  \left[  h_{l}^{\left(  2\right)  }\left(
k_{0}\left\vert \mathbf{x}\right\vert \right)  +e^{2i\delta_{l}}h_{l}^{\left(
1\right)  }\left(  k_{0}\left\vert \mathbf{x}\right\vert \right)  \right]
P_{l}\left(  \cos\theta\right)  ,\label{555}%
\end{equation}
where $k_{0}=\omega/c_{0}$, $c_{0}$ is the velocity of the acoustic wave
outside the nonrigid sphere, and $\rho_{0}$ is the density outside the
nonrigid sphere.

The phase shift $\delta_{l}$ is given by the connection condition on the
boundary \cite{schroeder2007handbook,morse1968theoretical}
\begin{equation}
\left.  u_{e}\left(  \mathbf{x}\right)  \right\vert _{r=a}=\left.
u_{o}\left(  \mathbf{x}\right)  \right\vert _{r=a}, \label{666}%
\end{equation}
and
\begin{equation}
\left.  \frac{1}{i\omega\rho_{\varepsilon}}\frac{\partial}{\partial r}%
u_{e}\left(  \mathbf{x}\right)  \right\vert _{r=a}=\left.  \frac{1}%
{i\omega\rho_{0}}\frac{\partial}{\partial r}u_{o}\left(  \mathbf{x}\right)
\right\vert _{r=a}. \label{777}%
\end{equation}
Substituting Eqs. (\ref{444}) and (\ref{555}) into Eqs. (\ref{666}) and
(\ref{777}) respectively gives
\begin{align}
e^{2i\delta_{l}}  &  =\left.  \frac{\frac{\rho_{e}}{\rho_{0}}j_{l}\left(
k_{e}r\right)  \frac{dh_{l}^{\left(  2\right)  }\left(  k_{0}r\right)  }%
{dr}-h_{l}^{\left(  2\right)  }\left(  k_{0}a\right)  \frac{dj_{l}\left(
k_{e}r\right)  }{dr}}{h_{l}^{\left(  1\right)  }\left(  k_{0}r\right)
\frac{d}{dr}j_{l}\left(  k_{e}r\right)  -\frac{\rho_{e}}{\rho_{0}}j_{l}\left(
k_{e}r\right)  \frac{dh_{l}^{\left(  1\right)  }\left(  k_{0}r\right)  }{dr}%
}\right\vert _{r=a}\nonumber\\
&  =\frac{-j_{l}\left(  k_{e}a\right)  \left[  l\left(  \rho_{0}-\rho
_{e}\right)  h_{l}^{\left(  2\right)  }\left(  k_{0}a\right)  +ak_{0}\rho
_{e}h_{l+1}^{\left(  2\right)  }\left(  k_{0}a\right)  \right]  +ak_{e}%
\rho_{0}h_{l}^{\left(  2\right)  }\left(  k_{0}a\right)  j_{l+1}\left(
k_{e}a\right)  }{j_{l}\left(  k_{e}a\right)  \left[  l\left(  \rho_{0}%
-\rho_{e}\right)  h_{l}^{\left(  1\right)  }\left(  k_{0}a\right)  +ak_{0}%
\rho_{e}h_{l+1}^{\left(  1\right)  }\left(  k_{0}a\right)  \right]
-ak_{e}\rho_{0}h_{l}^{\left(  1\right)  }\left(  k_{0}a\right)  j_{l+1}\left(
k_{e}a\right)  }, \label{e2id}%
\end{align}
where $\frac{d}{dz}h_{\nu}^{\left(  1,2\right)  }\left(  z\right)  =\frac{\nu
}{z}h_{\nu}^{\left(  1,2\right)  }\left(  z\right)  -h_{\nu+1}^{\left(
1,2\right)  }\left(  z\right)  $ and $\frac{d}{dz}j_{\nu}\left(  z\right)
=\frac{\nu}{z}j_{\nu}\left(  z\right)  -j_{\nu}\left(  z\right)  $ are used
\cite{olver2010nist}. The phase shift then reads
\begin{equation}
\delta_{l}=\arg\left(  j_{l}\left(  k_{e}a\right)  \left[  l\left(  \rho
_{0}-\rho_{e}\right)  h_{l}^{\left(  2\right)  }\left(  k_{0}a\right)
+ak_{0}\rho_{e}h_{l+1}^{\left(  2\right)  }\left(  k_{0}a\right)  \right]
-ak_{e}\rho_{0}h_{l}^{\left(  2\right)  }\left(  k_{0}a\right)  j_{l+1}\left(
k_{e}a\right)  \right)  -\frac{\pi}{2}. \label{787878}%
\end{equation}

Substituting Eq. (\ref{e2id}) into Eq. (\ref{77}) gives the scattering
intensity,%
\begin{align}
I\left(  \mathbf{x}\right)   &  =\frac{\rho_{0}\omega k_{0}}{\left\vert
\mathbf{x}\right\vert ^{2}}\left\vert \sum_{l=0}^{\infty}\left(  2l+1\right)
\frac{1}{2k_{0}}P_{l}\left(  \cos\theta\right)  y_{l}\left(  \frac{i}%
{k_{0}\left\vert \mathbf{x}\right\vert }\right)  \right. \nonumber\\
&  \times\left.  \left\{  \frac{2j_{l}\left(  k_{e}a\right)  \left[  l\left(
\rho_{0}-\rho_{e}\right)  j_{l}\left(  k_{0}a\right)  +ak_{0}\rho_{e}%
j_{l+1}\left(  k_{0}a\right)  \right]  -2ak_{e}\rho_{0}j_{l}\left(
k_{0}a\right)  j_{l+1}\left(  k_{e}a\right)  }{j_{l}\left(  k_{e}a\right)
\left[  l\left(  \rho_{0}-\rho_{e}\right)  h_{l}^{\left(  1\right)  }\left(
k_{0}a\right)  +ak_{0}\rho_{e}h_{l+1}^{\left(  1\right)  }\left(
k_{0}a\right)  \right]  -ak_{e}\rho_{0}h_{l}^{\left(  1\right)  }\left(
k_{0}a\right)  j_{l+1}\left(  k_{e}a\right)  }\right\}  \right\vert ^{2},
\label{200}%
\end{align}
where $j_{l}\left(  z\right)  =\frac{1}{2}\left[  h_{l}^{\left(  2\right)
}\left(  z\right)  +h_{l}^{\left(  1\right)  }\left(  z\right)  \right]  $ is
used \cite{liu2014scattering}.

\textit{Rayleigh scattering.}\textbf{ }Taking only the $s$-wave and $p$-wave
contributions into account and assuming that $ka\ll1$ gives the scattering
intensity
\begin{align}
I\left(  \mathbf{x}\right)   &  \sim\frac{\rho_{0}\omega k_{0}^{5}a^{6}%
}{9\left\vert \mathbf{x}\right\vert ^{2}}\left(  \frac{\kappa_{0}-\kappa_{e}%
}{\kappa_{0}}-3\frac{\rho_{e}-\rho_{0}}{2\rho_{e}+\rho_{0}}\cos\theta\right)
^{2}\nonumber\\
&  \times\left[  1+\left(  \frac{2\rho_{e}+\rho_{0}}{\rho_{e}-\rho_{0}}%
\frac{\kappa_{0}-\kappa_{e}}{3\kappa_{0}}\frac{1}{\cos\theta}-1\right)
^{-2}\frac{1}{k_{0}^{2}\left\vert \mathbf{x}\right\vert ^{2}}\right]  ,
\label{203}%
\end{align}
where $\kappa_{0}=k_{0}^{2}/\rho_{0}$ and $\kappa_{e}=k_{e}^{2}/\rho_{e}$.

When $\left\vert \mathbf{x}\right\vert $ tends to infinity, Eq. (\ref{203})
reduces to
\begin{equation}
I_{\infty}\left(  \mathbf{x}\right)  \sim\frac{\rho_{0}\omega k_{0}^{5}a^{6}%
}{9\left\vert \mathbf{x}\right\vert ^{2}}\left[  \frac{\kappa_{0}-\kappa_{e}%
}{\kappa_{0}}-\frac{3\left(  \rho_{e}-\rho_{0}\right)  }{\rho_{0}+2\rho_{e}%
}\cos\theta\right]  ^{2}. \label{104}%
\end{equation}
This is just the scattering intensity given by conventional scattering theory
\cite{schroeder2007handbook,morse1968theoretical}.

\subsection{Near target effect}

In the following, we show the influence of the near target effect in
scattering cross sections. Substituting $I\left(  \mathbf{x}\right)  $ given
by Eq. (\ref{200}) into Eq. (\ref{jiemian}) gives the differential scattering
cross section. In figures (\ref{nonka_0}) to (\ref{nonka_4}), we sketch
$\left(  \frac{d\sigma}{d\Omega}\right)  ^{1/2}/a$ versus the polar angel
$\theta$ at different distance $\left\vert x\right\vert $ for different wave
lengths up to $f$-wave contribution.

{\begin{figure}[ptb]
\centering
\includegraphics[width=0.8\textwidth]{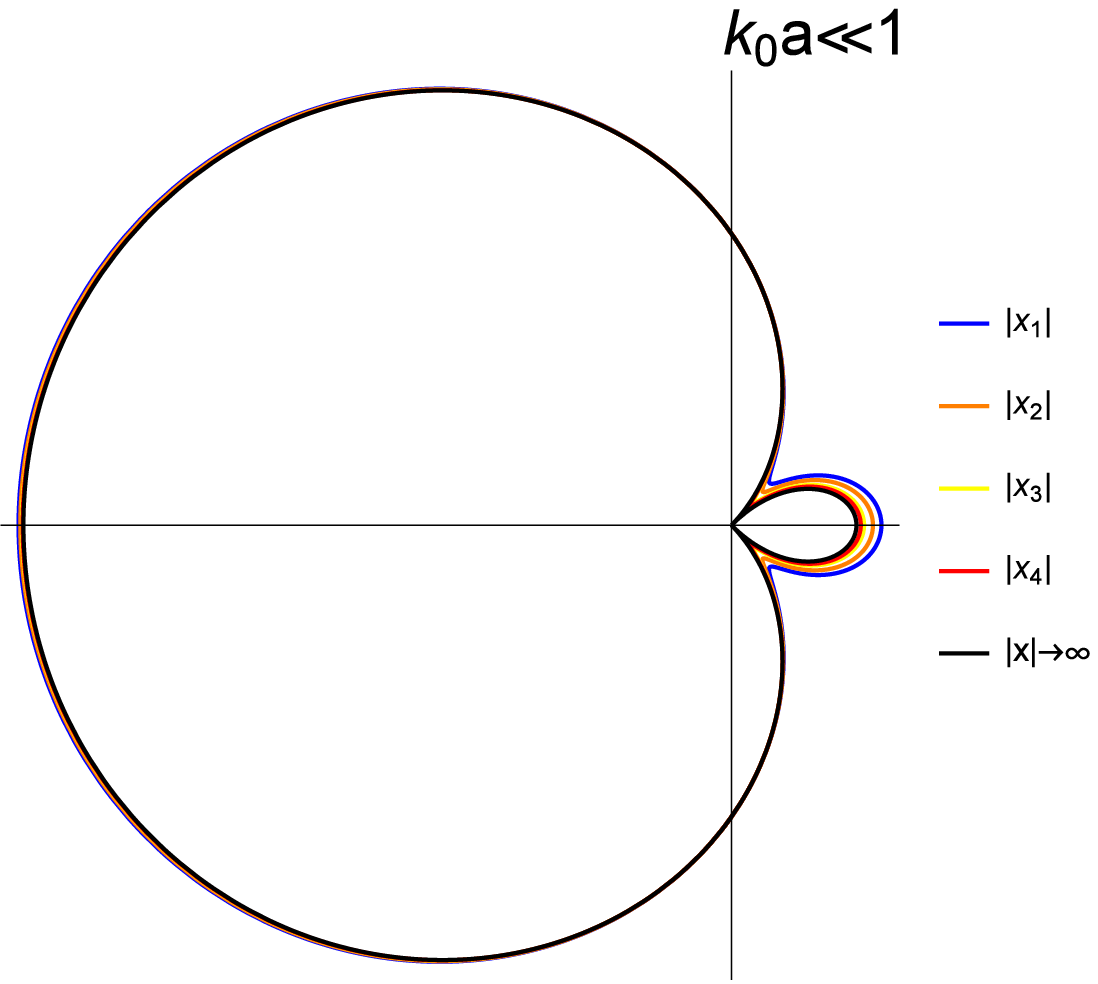}\newline\caption{A
sketch of the angular distribution of a plane wave scattered by a rigid sphere
of radius $a$: $\left(  \frac{d\sigma}{d\Omega}\right)  ^{1/2}/a$ versus the
polar angle $\theta$ for $k_{0}a\ll1$ up to the $f$-wave contribution at
different distance $\left\vert x_{1}\right\vert <\left\vert x_{2}\right\vert
<\left\vert x_{3}\right\vert <\left\vert x_{4}\right\vert <\left\vert
x\right\vert $.}%
\label{nonka_0}%
\end{figure}}

{\begin{figure}[ptb]
\centering
\includegraphics[width=0.8\textwidth]{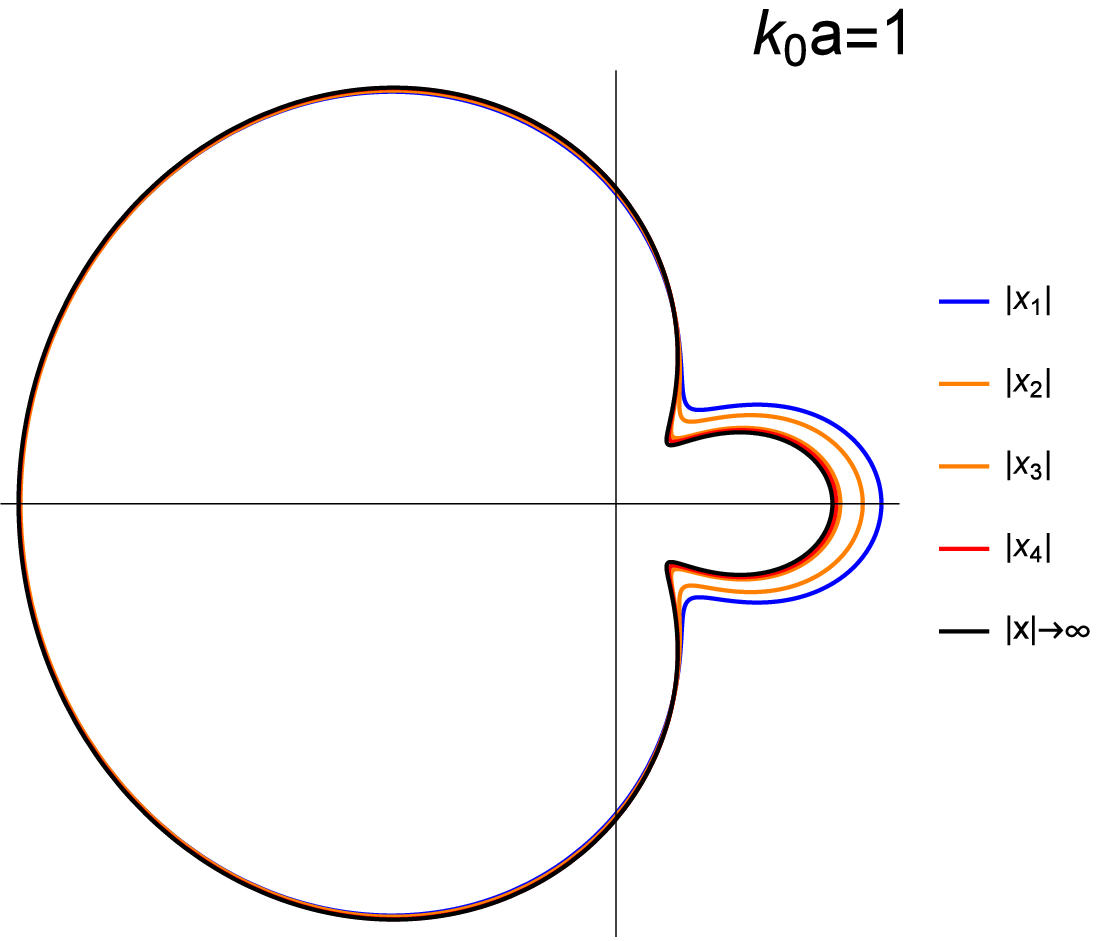}\newline\caption{A
sketch of the angular distribution of a plane wave scattered by a rigid sphere
of radius $a$: $\left(  \frac{d\sigma}{d\Omega}\right)  ^{1/2}/a$ versus the
polar angle $\theta$ for $k_{0}a=1$ up to the $f$-wave contribution at
different distance $\left\vert x_{1}\right\vert <\left\vert x_{2}\right\vert
<\left\vert x_{3}\right\vert <\left\vert x_{4}\right\vert <\left\vert
x\right\vert $.}%
\label{nonka_1}%
\end{figure}}

{\begin{figure}[ptb]
\centering
\includegraphics[width=0.8\textwidth]{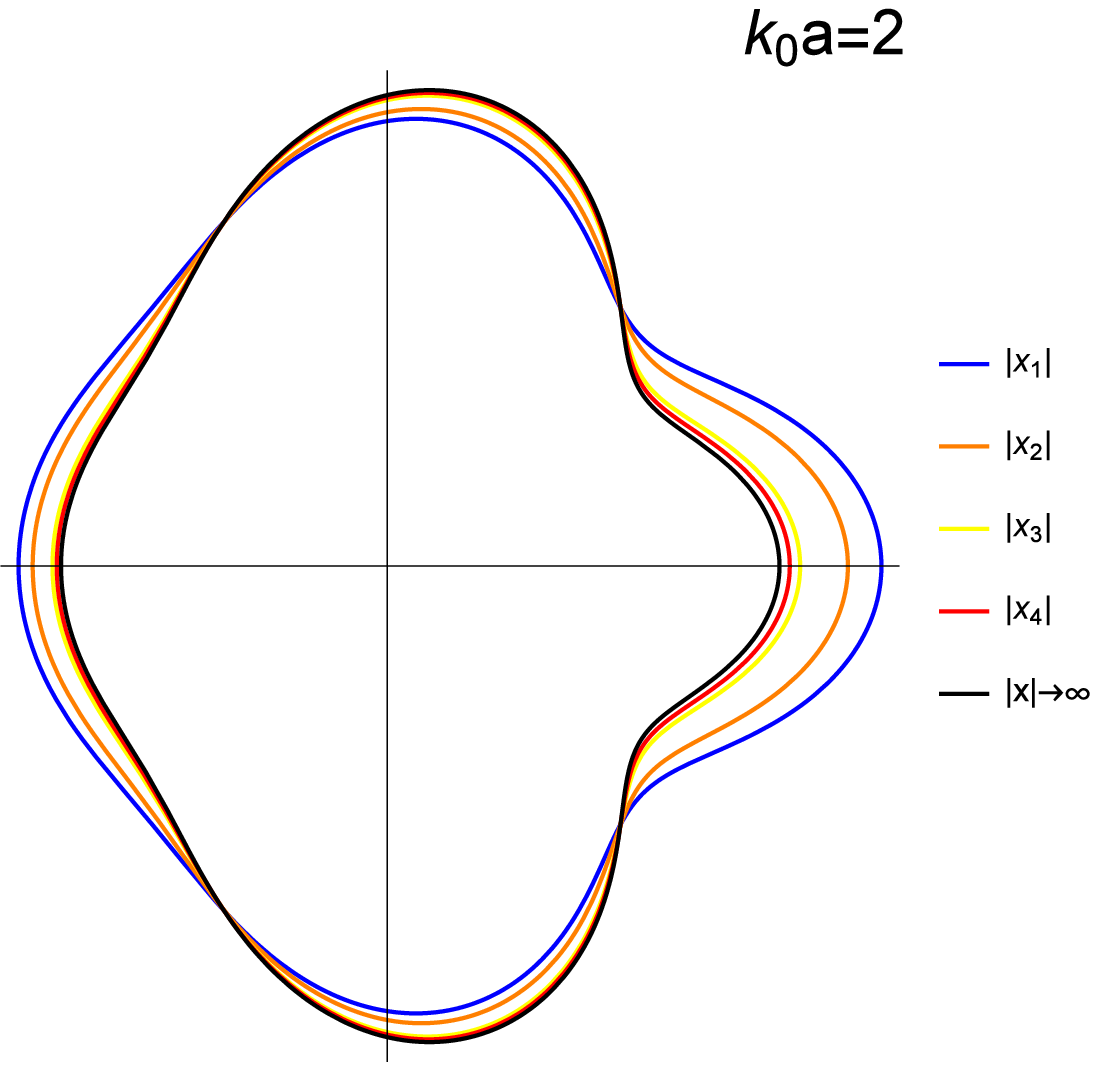}\newline\caption{A
sketch of the angular distribution of a plane wave scattered by a rigid sphere
of radius $a$: $\left(  \frac{d\sigma}{d\Omega}\right)  ^{1/2}/a$ versus the
polar angle $\theta$ for $k_{0}a=2$ up to the $f$-wave contribution at
different distance $\left\vert x_{1}\right\vert <\left\vert x_{2}\right\vert
<\left\vert x_{3}\right\vert <\left\vert x_{4}\right\vert <\left\vert
x\right\vert $. }%
\label{nonka_2}%
\end{figure}}

{\begin{figure}[ptb]
\centering
\includegraphics[width=0.8\textwidth]{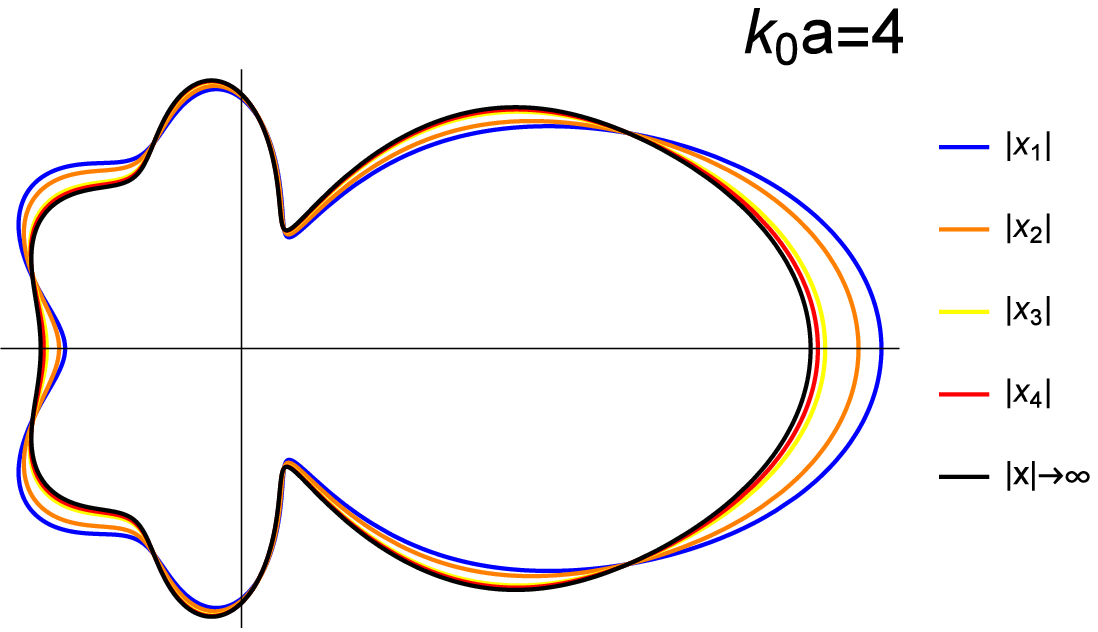}\newline\caption{A
sketch of the angular distribution of a plane wave scattered by a rigid sphere
of radius $a$: $\left(  \frac{d\sigma}{d\Omega}\right)  ^{1/2}/a$ versus the
polar angle $\theta$ for $k_{0}a=4$ up to the $f$-wave contribution at
different distance $\left\vert x_{1}\right\vert <\left\vert x_{2}\right\vert
<\left\vert x_{3}\right\vert <\left\vert x_{4}\right\vert <\left\vert
x\right\vert $.}%
\label{nonka_4}%
\end{figure}}

In order to show the near target effect, we consider the deviation
$\Delta\sigma=\sigma-\sigma_{\infty}$. By Eqs. (\ref{jiemian}), (\ref{203}),
and (\ref{104}), we have%
\begin{equation}
\frac{d\Delta\sigma}{d\Omega}=\frac{k_{0}^{2}a^{6}}{9\left\vert \mathbf{x}%
\right\vert ^{2}}\left(  \frac{2\rho_{e}+\rho_{0}}{\rho_{e}-\rho_{0}}%
\frac{\kappa_{0}-\kappa_{e}}{3\kappa_{0}}\frac{1}{\cos\theta}-1\right)  ^{-2}.
\end{equation}
where we take only the $s$-wave and $p$-wave contributions into account and
assume that $ka\ll1$.

\subsection{Power}

The scattering power can be obtained by substituting Eq. (\ref{203})\ into Eq.
(\ref{10222}).
\begin{equation}
P_{sc}(\omega,\mathbf{x})\sim\frac{4\pi}{9}\rho_{0}\omega k_{0}^{5}%
a^{6}\left[  \frac{\left(  \kappa_{0}-\kappa_{e}\right)  ^{2}}{\kappa_{0}^{2}%
}+3\left(  \frac{\rho_{e}-\rho_{0}}{2\rho_{e}+\rho_{0}}\right)  ^{2}\left(
1+\frac{1}{k^{2}\left\vert \mathbf{x}\right\vert ^{2}}\right)  \right]  .
\label{105}%
\end{equation}

When $\left\vert \mathbf{x}\right\vert $ tends to infinity, Eq. (\ref{105})
reduces to
\begin{equation}
P_{sc\infty}(\omega)\sim\frac{4\pi}{9}\rho_{0}\omega k_{0}^{5}a^{6}\left[
\frac{\left(  \kappa_{0}-\kappa_{e}\right)  ^{2}}{\kappa_{0}^{2}}+3\left(
\frac{\rho_{e}-\rho_{0}}{2\rho_{e}+\rho_{0}}\right)  ^{2}\right]  .
\label{106}%
\end{equation}
This is just the scattering power given by the conventional scattering theory
\cite{schroeder2007handbook,morse1968theoretical}.

\section{Discussions and conclusions \label{Conclusion}}

In this paper, we provide a rigorous acoustic scattering theory without the
large-distance asymptotic approximation. In our treatment, instead of the
far-field pattern $u_{\infty}\left(  \frac{\mathbf{x}}{\left\vert
\mathbf{x}\right\vert }\right)  $ in conventional scattering theory, we
consider an exact pattern $u_{exact}\left(  \mathbf{x}\right)  $ without the
large-distance asymptotic approximation. The scattering intensity $I\left(
\mathbf{x}\right)  $ is represented in terms of the exact pattern
$u_{exact}\left(  \mathbf{x}\right)  $. The exact pattern $u_{exact}\left(
\mathbf{x}\right)  $ is represented in terms of the phase shift $\delta_{l}$.
Two examples, plane waves scattering on a rigid sphere and on a nonrigid
sphere are considered. We also compare our result with the result given by
conventional scattering theory. Moreover, we show that the near target effect
can not be ignored for long wavelength acoustic scattering.

For an incident wave packet $u^{inc}\left(  \mathbf{r}\right)  $ along the
$z$-axis, the incident wave can be expanded in terms of plane waves
$e^{i\mathbf{k}\cdot\mathbf{r}}$ with $\mathbf{k}$ the wave vector in $z$
direction, that is, $u^{inc}\left(  \mathbf{r}\right)  =\sum_{\mathbf{k}%
}C_{\mathbf{k}}e^{i\mathbf{k}\cdot\mathbf{r}}$, where $C_{\mathbf{k}}$ is the
coefficient. Each incident plane wave $e^{i\mathbf{k}\cdot\mathbf{r}}$
corresponds to a scattering wave $u_{\mathbf{k}}^{sc}\left(  \mathbf{r}%
\right)  $. Therefore, the scattering wave $u^{sc}\left(  \mathbf{r}\right)  $
of incident wave packet $u^{inc}\left(  \mathbf{r}\right)  $ is the sum of
scattering waves $u_{\mathbf{k}}^{sc}\left(  \mathbf{r}\right)  $ of each
individual incident plane waves $e^{i\mathbf{k}\cdot\mathbf{r}}$, i.e.,
$u^{sc}\left(  \mathbf{r}\right)  =\sum_{\mathbf{k}}C_{\mathbf{k}%
}u_{\mathbf{k}}^{sc}\left(  \mathbf{r}\right)  $. That is, the key problem is
the plane-wave scattering, so in this paper we only\ consider the case that
the incident wave is a plane wave. Moreover, when the incident wave is a wave
packet, one also need to consider the contribution of the interference.

The scattering considered in this paper is the scattering on spherically
symmetric target. For non-spherically symmetric targets, the target function
can be expanded by the spherical harmonic function \cite{boardman1967partial}.
Then the result of a non-spherically symmetric scattering is a series sum of
spherically symmetric scatterings.

In realistic acoustic scattering, the wavelength of acoustic waves is often
comparable with the distance between the target and the observer. In this
case, a rigorous theory of acoustic scattering that preserves the information
of the distance becomes important.

In acoustic scattering, it is needed to\ calculate the scattering phase shift.
The scattering phase shift is determined by solving the Helmholtz equation
with the boundary condition on the obstacle. In further works, we can apply
the heat-kernel method to calculate the scattering phase shift using the
approach developed in Refs.
\cite{graham2009spectral,barvinsky1987beyond,pang2012relation,li2015heat}.

In long wavelength scattering, such as the scattering of sounds with
wavelength ranging from tens of centimeters to a few meters, sonars with
wavelength ranging from tens of meters to hundreds of meters, and earthquake
waves with wavelength ranging from hundreds of meters to thousands of meters,
the wavelength of acoustic waves is often too long to use the large-distance
asymptotic approximation. In applications such as designing the structure of
the theater, conducting deep sea and geological exploration by detecting the
sonar and the earthquake wave, the distance is comparable with the wavelength
in order to receive signals with enough intensity. Under this circumstance the
near target effect needs to be reckoned in.


\acknowledgments

We are very indebted to Dr G. Zeitrauman for his encouragement. This work is supported in part by NSF of China under Grant
No. 11575125 and No. 11675119.






\end{CJK*}
\end{document}